\documentclass[]{spie}  

\usepackage{amsmath,amsfonts,amssymb}
\usepackage{graphicx}
\usepackage[colorlinks=true, allcolors=blue]{hyperref}
\usepackage{caption}
\usepackage{subcaption}
\usepackage{dblfloatfix}

\title{Ghosts of NEID’s Past} 

\author[a,b]{Shubham Kanodia}
\author[a,b]{Joe P.\ Ninan}
\author[a,b]{Andrew J. Monson}
\author[a,b]{Suvrath Mahadevan}
\author[a, b]{Colin Nitroy}
\author[c]{Christian Schwab}
\author[d]{Samuel Halverson}
\author[e]{Chad F. Bender}
\author[f]{Ryan Terrien}
\author[a,b]{Frederick R. Hearty}
\author[g]{Emily Lubar}
\author[h]{Michael W. McElwain}
\author[a,b]{Lawrence. W. Ramsey}
\author[i]{Paul M. Robertson}
\author[j]{Arpita Roy}
\author[k]{Gudmundur Stefansson}
\author[a,b]{Daniel J. Stevens}

\affil[a]{Department of Astronomy \& Astrophysics, The Pennsylvania State University, 525 Davey Lab, University Park, PA 16802, USA}
\affil[b]{Center for Exoplanets \& Habitable Worlds, The Pennsylvania State University, University
Park, PA 16802, USA}
\affil[c]{Department of Physics and Astronomy, Macquarie University, Balaclava Road, North Ryde, NSW 2109, Australia}
\affil[d]{Jet Propulsion Laboratory, 4800 Oak Grove Drive, Pasadena, CA 91109, USA}
\affil[e]{Steward Observatory, The University of Arizona, 933 N.\ Cherry Ave, Tucson, AZ 85721, USA}
\affil[f]{Department of Physics and Astronomy, Carleton College, One North College Street, Northfield, MN 55057, USA}
\affil[g]{Department of Astronomy, University of Texas at Austin, Physics Mathematics Astronomy building, E Dean Keeton St \& Speedway, Austin, TX 78712}
\affil[h]{NASA’s Goddard Space Flight Center, Greenbelt, MD 20771, USA}
\affil[i]{Department of Physics \& Astronomy, University of California - Irvine, 4129 Frederick Reines Hall, Irvine, CA 92697, USA}
\affil[j]{Space Telescope Science Institute, 3700 San Martin Dr, Baltimore, MD 21218, USA}
\affil[k]{Princeton University, Princeton, NJ 08540, USA}

\authorinfo{Further author information: E-mail: szk381@psu.edu}

\newcommand\cms{$\textrm{cm~s}^{-1}$}

\begin{document} 
\maketitle

\begin{abstract}
The NEID spectrograph is a R $\sim$ 120,000 resolution fiber-fed and highly stabilized spectrograph for extreme radial velocity (RV) precision.  It is being commissioned at the 3.5 m WIYN telescope in Kitt Peak National Observatory with a desired instrumental precision of better than 30 \cms{}. NEID’s bandpass of 380 -- 930 nm enables the simultaneous wavelength coverage of activity indicators from the Ca HK lines in the blue to the Ca IR triplet in the IR. In this paper we will present our efforts to characterize and mitigate optical ghosts in the NEID spectrograph during assembly, integration and testing, and highlight several of the dominant optical element contributors such as the cross dispersion prism and input optics. We shall present simulations of the 2-D spectrum and discuss the predicted ghost features on the focal plane, and how they may impact the RV performance for NEID. We also present the mitigation strategy adopted for each ghost which may be applied to future instrument designs. This work will enable other instrument builders to potentially avoid some of these issues, as well as outline mitigation strategies.

\end{abstract}

\keywords{Spectrometer, Optical Design, Ghost analysis, Scattered light}

\section{INTRODUCTION}\label{sec:intro}  
The radial velocity (RV) method of planet detection was used to detect the first exoplanet (a hot Jupiter) orbiting a solar type star in 1995 - 51 Pegasi b \cite{mayor_jupiter-mass_1995}. Since then the instrumentation and reduction techniques have improved vastly; with the new generation of extreme precision RV spectrographs moving towards the precision required for detecting Earth analogues with semi-amplitudes of $\sim 9$ \cms{}. Such precise RV measurements are needed for measuring the masses of planet candidates from transiting surveys, as well as for discovering terrestrial mass exoplanets. In addition, these measurements are required for atmospheric characterization using the upcoming James Webb Space Telescope\cite{batalha_precision_2019}. The next generation of flagship space-based direct imaging missions such as  LUVOIR\cite{the_luvoir_team_luvoir_2019}, and HabEx\cite{gaudi_habitable_2019} also benefit from the determination of orbital parameters for optimizing their instrument and survey strategies.  The new generation of precise RV spectrographs have extensive environmental stability \cite{mayor_setting_2003, robertson_system_2016, stefansson_versatile_2016}, are generally fiber illuminated, and to economize the size of the optical elements have gravitated towards the white pupil optical design \cite{baranne_white_1988}. 

\section{NEID}
\subsection{Overview}\label{sec:neidoverview}
NEID is a next-generation precision RV instrument being commissioned at the 3.5 m WIYN telescope at Kitt Peak, USA. The instrumental precision goal of NEID ($< 50$ \cms{}) necessitates stringent requirements on instrumental subsystems in the form of an instrumental (+telescope) error budget of 27 \cms{}. It has extreme environment control \cite{robertson_system_2016, stefansson_versatile_2016, robertson_ultrastable_2019} and is illuminated using optical fibers \cite{kanodia_overview_2018}. NEID has two operating modes: a High Resolution (HR) mode with a 62 $\mu$m fiber for maximum RV precision for use on brighter targets; and a High Efficiency (HE) mode for fainter targets with a bigger fiber ($\sim 100 \mu$m). The HR mode consists of 3 fibers - Science, Calibration and Sky, whereas the HE mode uses a Science and Sky fiber. The Sky fiber enables simultaneous measurement of sky background contamination, especially OH$^-$ lines, whereas the Calibration fiber in the HR mode allows for correcting instrumental drift by using a stable reference source such as a laser frequency comb (LFC) concurrent with the science exposure.

\begin{figure}[] 
\center
\includegraphics[width=0.75\columnwidth]
{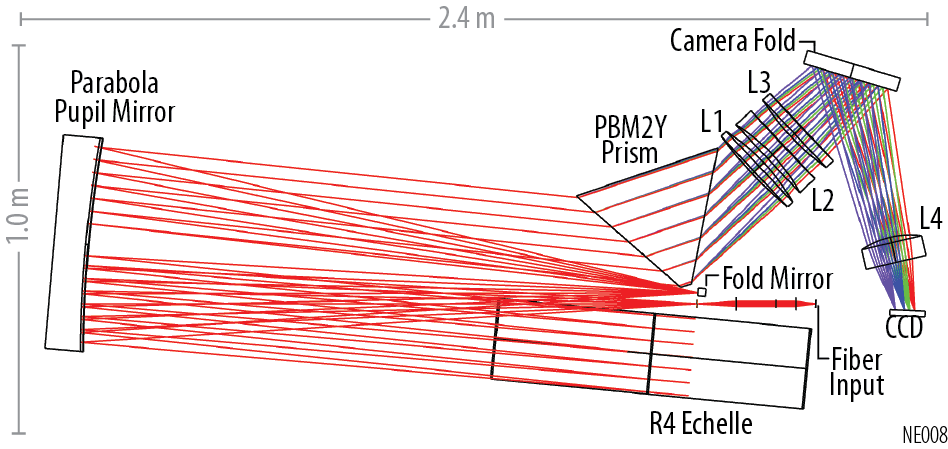}
\caption{White pupil  design for NEID using a mosaic R4 echelle and a single PBM2Y prism as the dispersive elements\cite{schwab_design_2016}.} \label{fig:opticaldesign}
\end{figure}

NEID has a wavelength coverage of 380 -- 930 nm, and a spectral resolution of 120,000. To achieve instrumental point spread function (PSF) stability, it uses a single fiber to couple light from the telescope to the instrument without any image or pupil slicing. With a symmetric white pupil design\cite{baranne_white_1988}, NEID uses an R4 echelle grating along with a prism cross-dispersing element since it's wavelength range spans more than an octave ($\lambda_{max} > 2\lambda_{min}$)\cite{schwab_design_2016} (\autoref{fig:opticaldesign}).

\subsection{NEID Optical Design}\label{sec:neidoptics}
The 5 instrument fibers (3 HR + 2 HE) are terminated in a fused silica fiber puck, after which a focal ratio converting system is used to convert the f/3.65 fiber output to f/8 using two cemented doublet lenses and a singlet lens\footnote{Hereafter referred to as the input optics.} To minimize Fresnel losses at the fiber output, we glue a plano-convex lens at the face of the puck using an index matching low expansion epoxy (\textit{Epotek} 301-2), which also slows the fiber output light cone to f/7.9. This lens along with the double doublet lenses form an intermediate focus which is then imaged on to the fold mirror (M2) by the parabola (M1) after being dispersed by the R4 echelle grating with a $\gamma$ angle of 0.5$^{\circ}$. The flat fold mirror reflects the beam on to the parabola for a third bounce, which then forms a white pupil on the incident face of PBM2Y equilateral prism. The PBM2Y prism ensures high efficiency across the NEID bandpass, including the UV ($< 400 $ nm). After the prism, the collimated beam is imaged on to the focal plane by a four element camera which includes a fold mirror for compactness. 

The optical elements are extensively baffled to minimize stray light contribution on the focal plane array (FPA), where the baffles are either coated with Z306 or carbon nanotubes to absorb the scattered light. In addition to this, we use \textit{OpticStudio} Zemax Non-Sequential mode to search for, and mitigate the presence of ghosts in NEID. In the next section we discuss the various ghosts discovered in NEID during the alignment and integration step.

\section{Ghosts seen in NEID}
For NEID we use the ``optimal extraction"\cite{baranne_elodie:_1996} technique to extract 1-D stellar spectra from the 2-D echellogram as imaged on the detector. This requires an accurate estimate of the background flux levels across the detector, as well as high S/N flat fields to correct for the relative variations in pixel level sensitivity. We interpolate between the orders to estimate a global background, which is then subtracted from the 2-D image. This assumes that any changes in the  background are at a low frequency, and that the background does not contain any sharp features. Apart from precise RV measurement, stray light contamination can be detrimental to spectroscopic studies which require an accurate estimation of the continuum levels to measure the equivalent width of lines. The inter-order background consists of decaying wings (in the cross dispersion direction) from the stellar, sky and calibration fiber spectra, as well as a diffuse background. Some of the sources of this diffused background are - 

\begin{itemize}
\item Detector noise - This not only includes dark and read noise, but also effects such as the non-uniformities in pixel response. These can be typically be corrected with precise pixel level calibration. In addition, most CCDs are operated at cryogenic temperatures to minimize the dark noise contribution.
\item Instrument thermal background - This is not an issue for NEID since it operates in the optical wavelengths, where the contribution from ambient surroundings at its nominal temperature of 300 K is extremely small. However this can be a major source of background noise for infrared instruments.
\item Zeroth order echelle reflection - The blazed echelle grating reflects majority of the light in littrow configuration in the orders where the angle of incidence ($\alpha$) is equal to the angle of diffraction ($\beta$). However in addition to this, a small fraction of the light incident on the echelle is reflected to other orders, the majority of which is in the `zeroth' order which is equivalent to specular reflection of the grating surface. 
\item Periodic grating groove errors - Periodic errors in the groove spacing of gratings can create false lines or ghosts (known as a Rowland ghost\cite{quincke_optische_1872, rowland_xxxix_1893, king_note_1903,gale_rowland_1937}). These ghosts are typically symmetrically placed around the primary spectra and can contaminate neighbouring orders. 
\item Surface scattering and irregularity - One of the biggest sources of scattered light can be the \textit{roughness} of the optical elements. This is generally minimized by keeping the optics clean with minimal scratches, as well as strict specifications during fabrication and polishing. In addition to this, low order wavefront errors in optical elements can exacerbate the PSF aberrations which can degrade the PSF.  
\item Reflections off optical mounts - The light from faster f/$\#$ (due to FRD) can reflect off the mounts or baffles (especially from the edges), and form sharp patterns on the detector. We minimize this for NEID by forming knife edges on the baffles to reduce back scatter. We also anodize the baffle surfaces or coat them with carbon nanotubes to absorb ghost and scattered beams at glancing incidence.
\item Optical Ghosts - Optical ghosts are typically formed from refractive elements and can be minimized with baffles and broad band anti-reflection coatings (BBAR). In subsequent sections, we discuss some of the ghosts seen in NEID, and the mitigation strategies employed therewith.
\end{itemize}

An optical ghost can be defined as a shape or feature present in the focal (image) plane of an optical instrument that is not present in the object plane. In the case of NEID, the object plane is represented as the fiber face which is imaged on to the detector by the optics of the system after dispersing (by the echelle grating) and cross-dispersing (by the prism). All refractive elements inside NEID are coated with an AR coating (which reflects $\le 1\%$ of the light) to minimize back reflections and increase throughput. Despite this, reflections at the 1$\%$ level can also have a negative impact when imaged on to the detector.  In subsequent sections we outline the ghosts we see in NEID, and discuss the solutions we attempted that worked (and also the ones that did not) to solve them. We do so, hoping to help future instruments to identify and mitigate ghosts and potential scattering mechanisms.

\begin{figure*}[!t]
  \centering
  \begin{tabular}[]{@{}p{0.39\textwidth}@{}}
  {
\includegraphics[width=0.38\textwidth]
{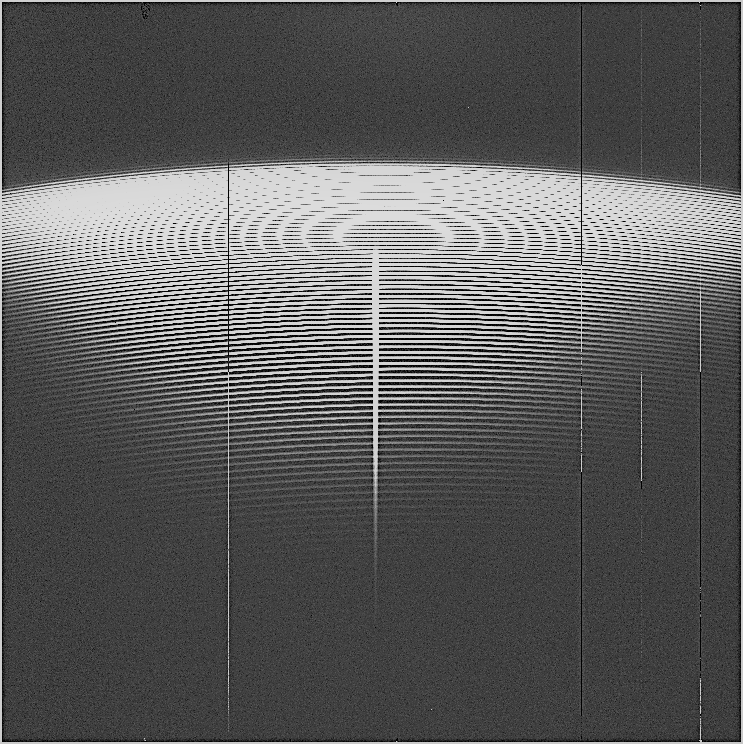}} \\\small (a) NEID FPA illuminated by a continuum source using a SMF. The bright streak in the middle is the prism apex ghost formed due to over-filling of the optics, whereas some of the bad columns in the engineering grade detector are also seen.
  \end{tabular}%
  \quad
  \begin{tabular}[]{@{}p{0.51\textwidth}@{}}
 {
 \includegraphics[width=0.5\textwidth]
{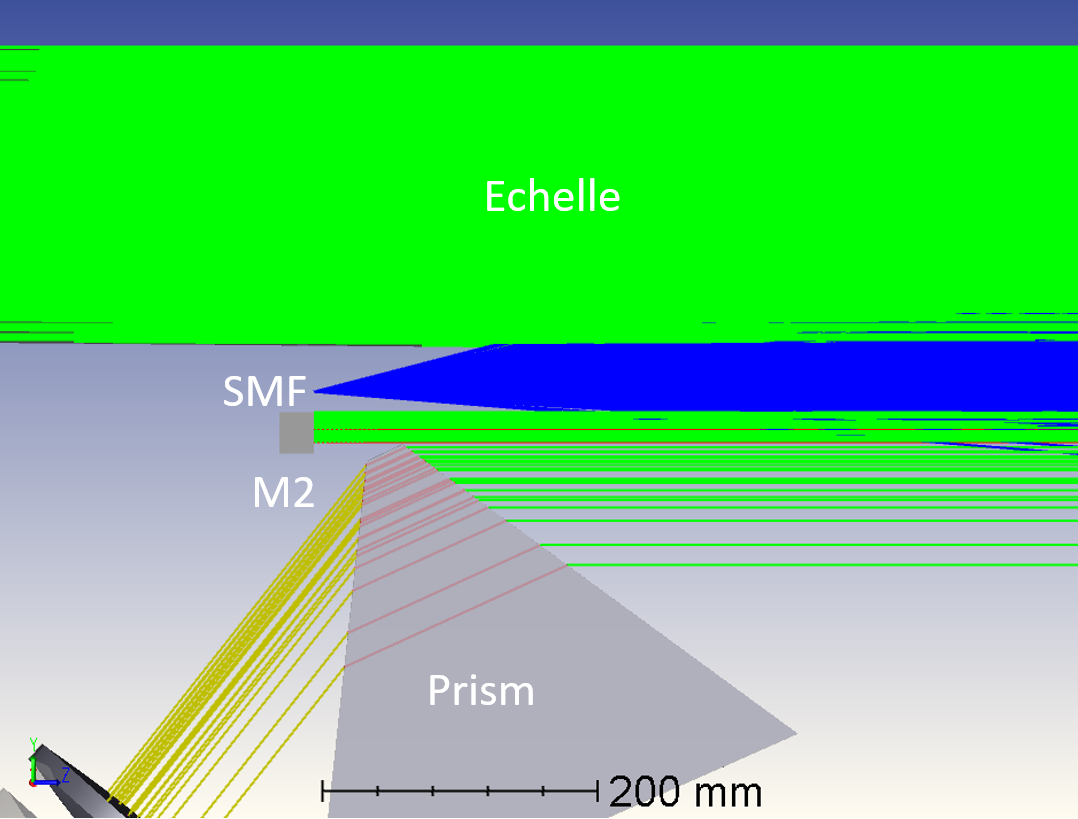}} \\
\small (b) \textit{Zemax} rendition of the SMF overfilling M1. The input beam from the SMF is shown in blue, whereas the return collimated beam from the parabola (M1) is shown in green. Since the SMF overfills the aperture, part of the reflected light skips the grating and transmits through the prism apex (yellow). 
  \end{tabular}
  \caption{The prism apex ghost }   \label{fig:prismapexghost}
\end{figure*}

\subsection{Prism apex ghost with SMF}\label{sec:prismapex}
During the first stage of the optical alignment process for NEID, we used a fused silica single mode fiber (SMF) with an output f/4 output at intermediate focus. The tiny (4.5 $\mu$m) fiber size allowed us to optimize the focus and minimize aberrations in the PSF for NEID. The SMF output of f/4 is faster than the nominal f/$\#$ expected by the baffle elements for the optics from intermediate focus\footnote{Expected f/$\#$ at intermediate focus is f/8.}, and hence overfills the parabola (M1). This reflected collimated beam which overfills M1 is directed towards the echelle grating, but also the fold mirror and the prism apex. The part of the beam which leaks and enters the prism apex forms a \textit{cross} dispersed streak on the detector that is only spread out in the cross dispersion direction and not the dispersion (since it misses the echelle grating). This streak is seen on the FPA (\autoref{fig:prismapexghost}a), whereas in \autoref{fig:prismapexghost}b we see a \textit{Zemax} simulation of this. To mitigate this ghost during alignment stage we added a temporary f/8 stop in front of the SMF which functioned as a baffle blocking the faster beam.

\begin{figure*}[!t]
  \centering
  \begin{tabular}[]{@{}p{0.75\textwidth}@{}}
  {
\includegraphics[width=0.62\textwidth]
{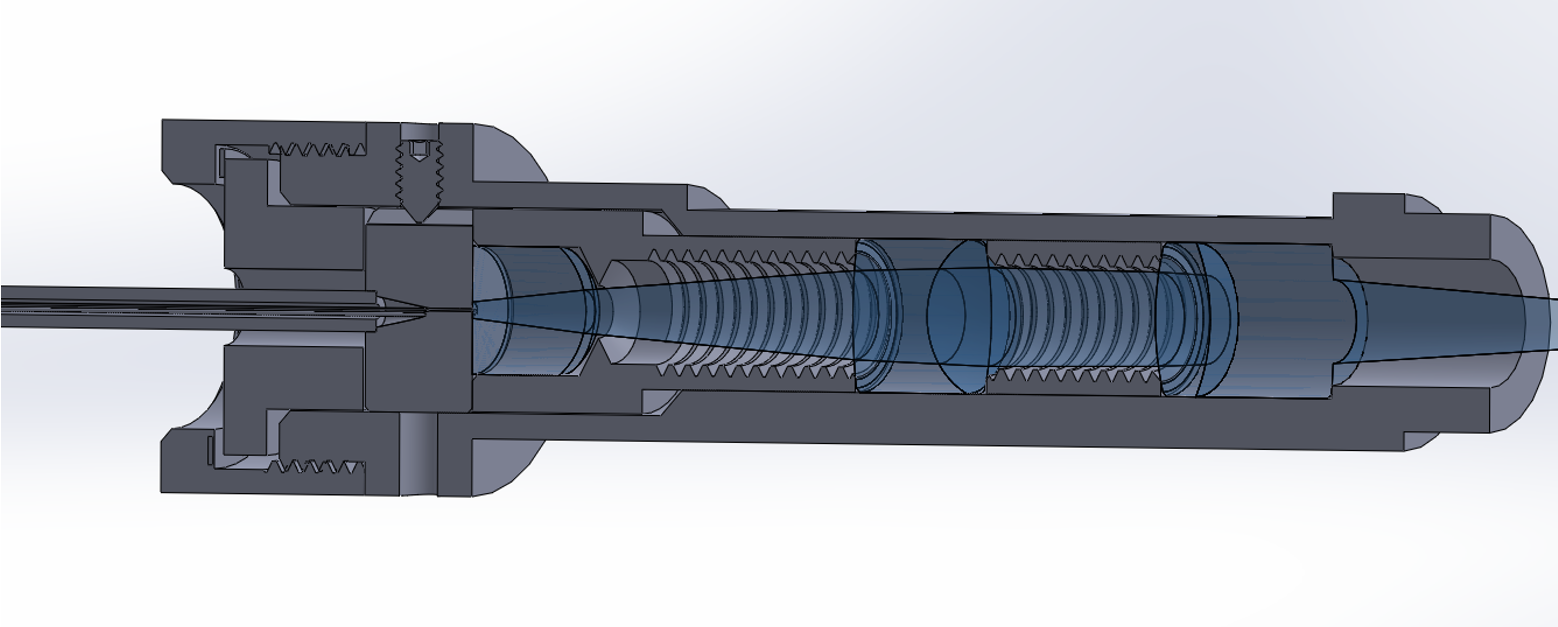}} \\
  \end{tabular}%
  \quad
  \caption{\textit{SolidWorks} rendition of the input fiber puck, and the optics tube (grey) showing the double doublet lenses (blue).}   \label{fig:inputopticstube}
\end{figure*}

\begin{figure}[!b]
\center
\includegraphics[width=0.4\columnwidth]
{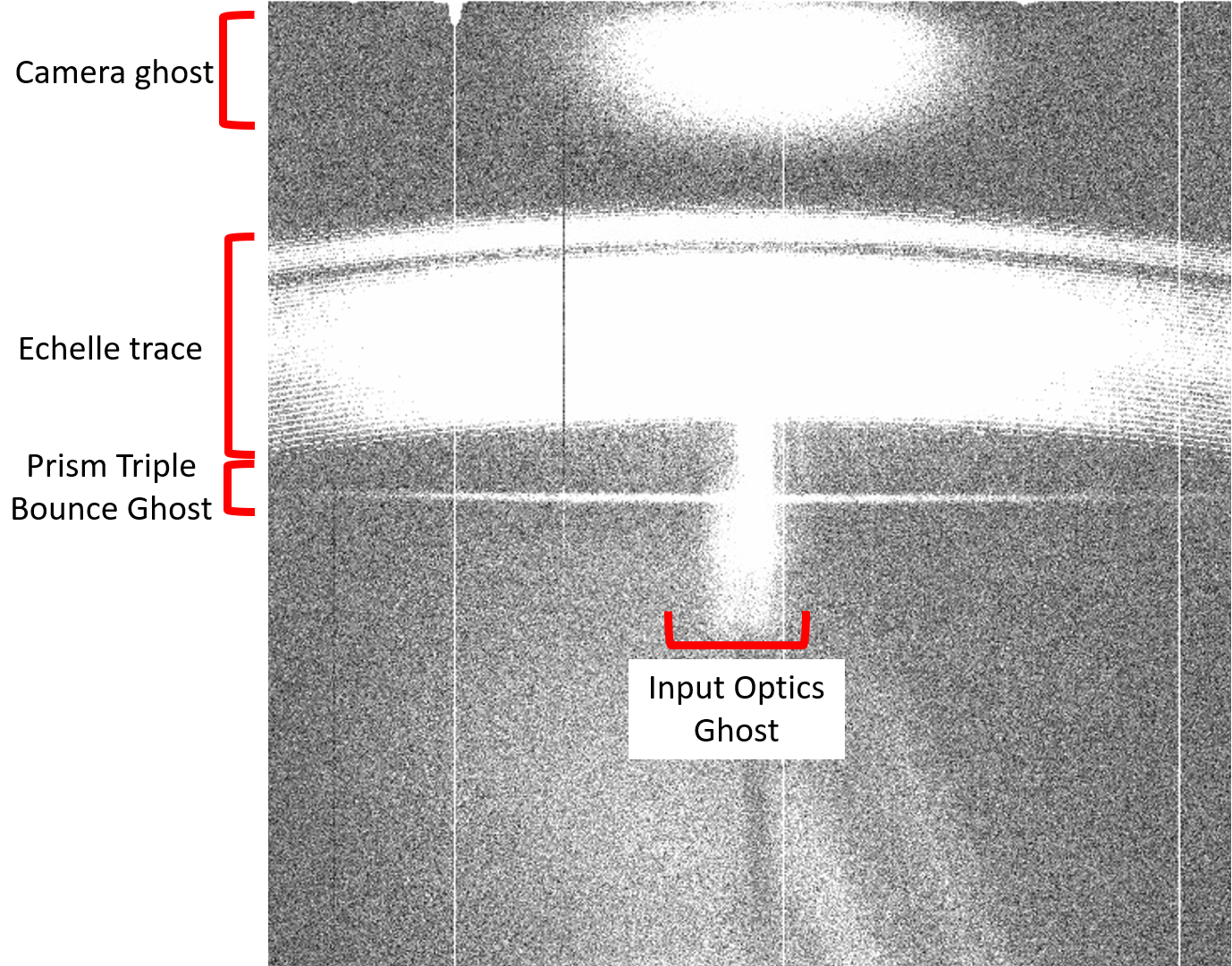}
\caption{Continuum illuminated echellogram on the NEID engineering detector. This image shows three of the ghosts discussed in this manuscript. The diffused patch on the top is the L3.2 ghost (Section \ref{sec:cameraghost}), followed by the central vertical streak from the input optics ghost (Section \ref{sec:inputopticsghost}). The sharp horizontal ghost is the prism triple bounce ghost (Section \ref{sec:prismtriple}). In addition, some of the bad pixel columns in the engineering grade detector as visible. } \label{fig:ThreeGhosts}
\end{figure}

\subsection{Input optics ghost}\label{sec:inputopticsghost}
After the initial alignment and focussing step with the SMF, we switched the input fiber to the final NEID fiber configuration using a fused silica puck with 5 fibers (3 HR + 2HE; Section \ref{sec:neidoverview}). The input optics consist of a singlet plano-convex lens and two doublet lenses to convert the f/3.65 fiber output to an f/8 beam. These lens are coated with a BBAR to minimize back reflections, and are mounted in an aluminium tube which is threaded on the inside, and also black anodized\footnote{ Anodize MIL-A-8625 Type I, Class 2 Black} to reduce scattered light\cite{leonhard_euler_letters_1802} (\autoref{fig:inputopticstube}). After switching to the input optics tube we noticed a central streak down order traces on the echellogram (\autoref{fig:ThreeGhosts}). Like the earlier prism apex ghost (Section \ref{sec:prismapex}), this ghost was also cross dispersed down the middle of the order, which implied that the ghost beam bypassed the echelle grating.

\begin{figure}[t]
\center
\includegraphics[width=0.95\columnwidth]
{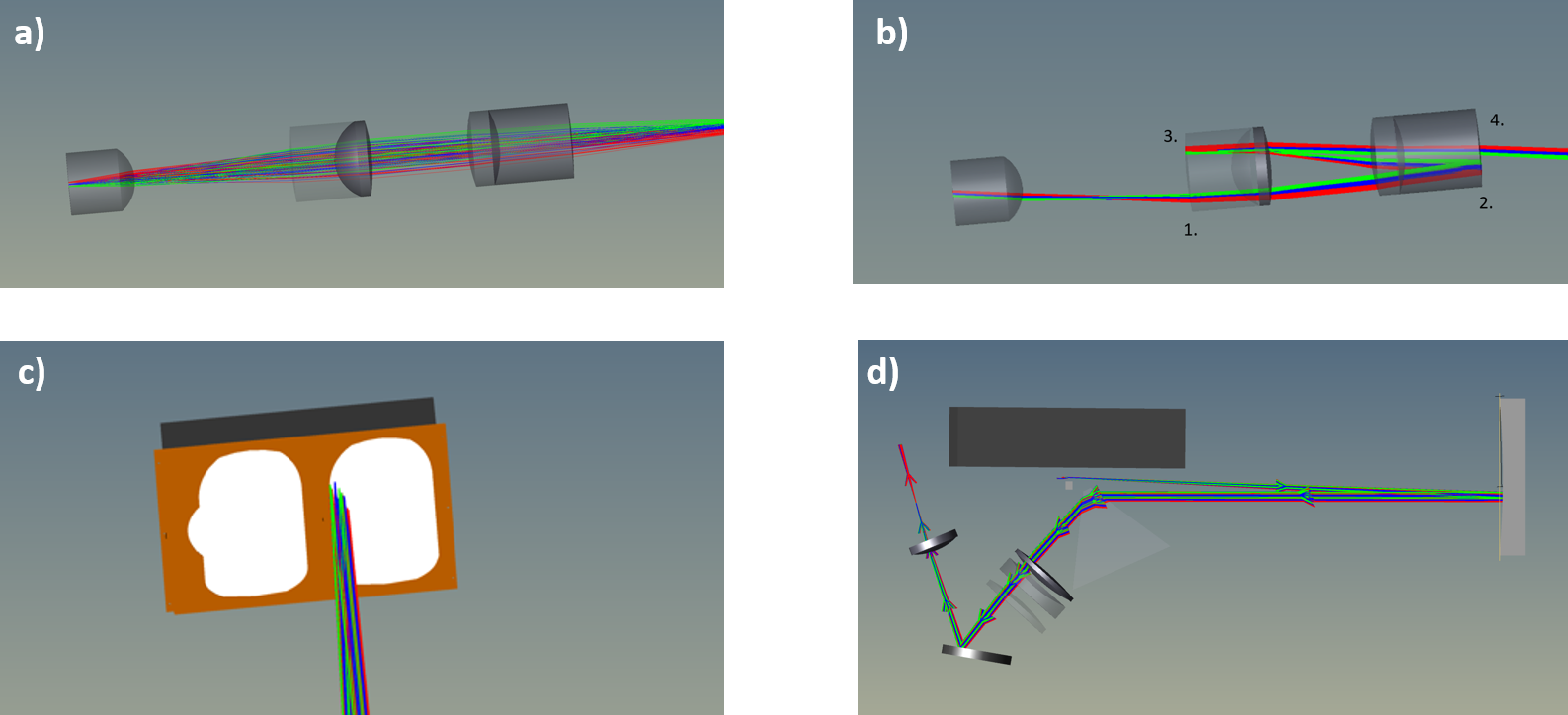}
\caption{\textit{Zemax} analysis done to investigate the source of the input optics ghost. \textbf{a)} The nominal f/3.65 science beam along the input optics lenses. \textbf{b)} The ghost reflections along the input optics. \textbf{c)} The ghost beam clips the edge of the M1 baffle which has been critically sized for the science beam for the 3 fibers. \textbf{d)} A top down view showing the ghost beam missing the grating, and getting dispersed by the prism on its way to the detector.} \label{fig:inputopticszemax}
\end{figure}

\begin{figure*}[!b]
  \centering
  \begin{tabular}[]{@{}p{0.45\textwidth}@{}}
  {
\includegraphics[width=0.42\textwidth]
{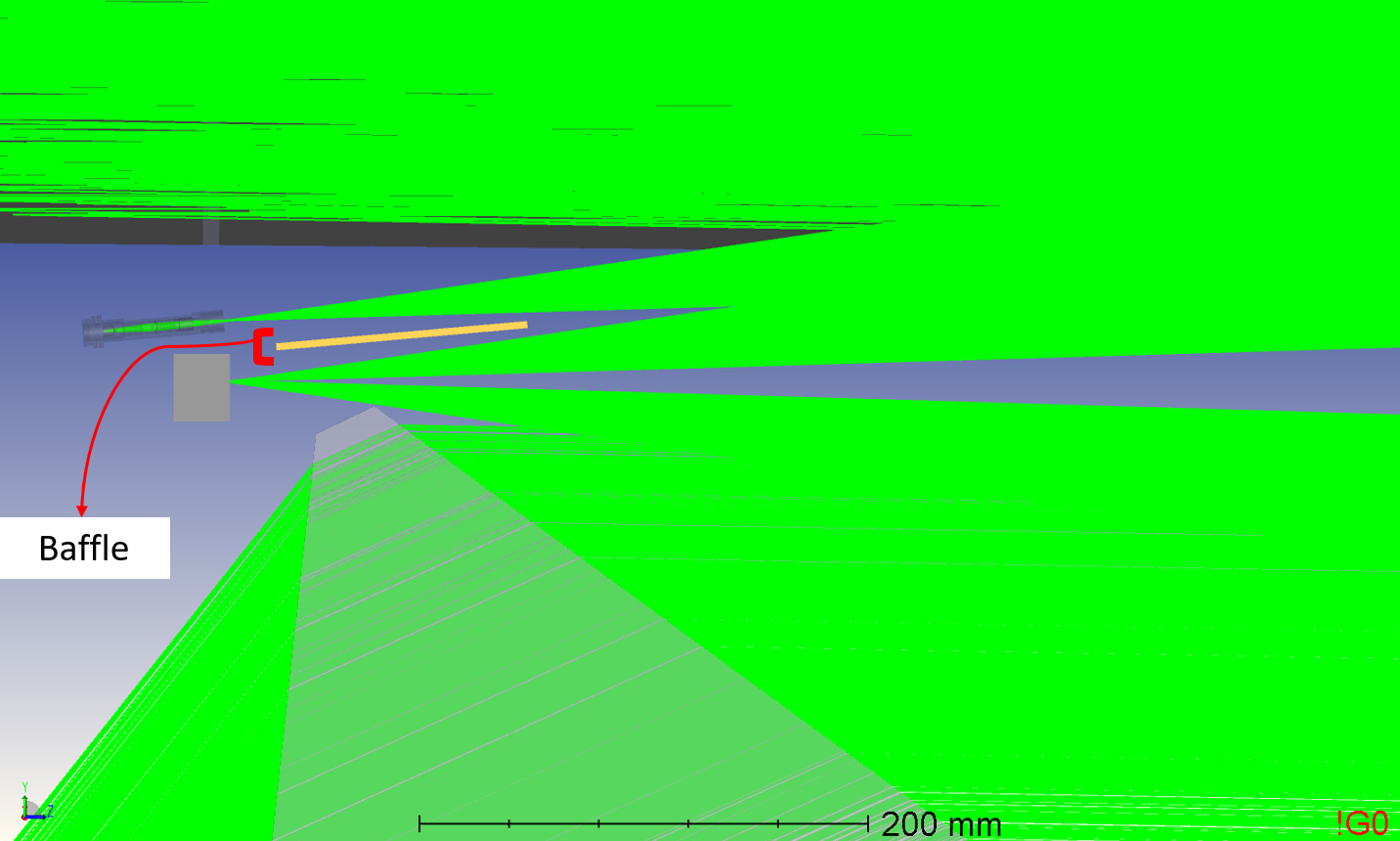}} \\\small (a) The science beam with respect to the baffle.
  \end{tabular}%
  \quad
  \begin{tabular}[]{@{}p{0.45\textwidth}@{}}
 {
 \includegraphics[width=0.42\textwidth]
{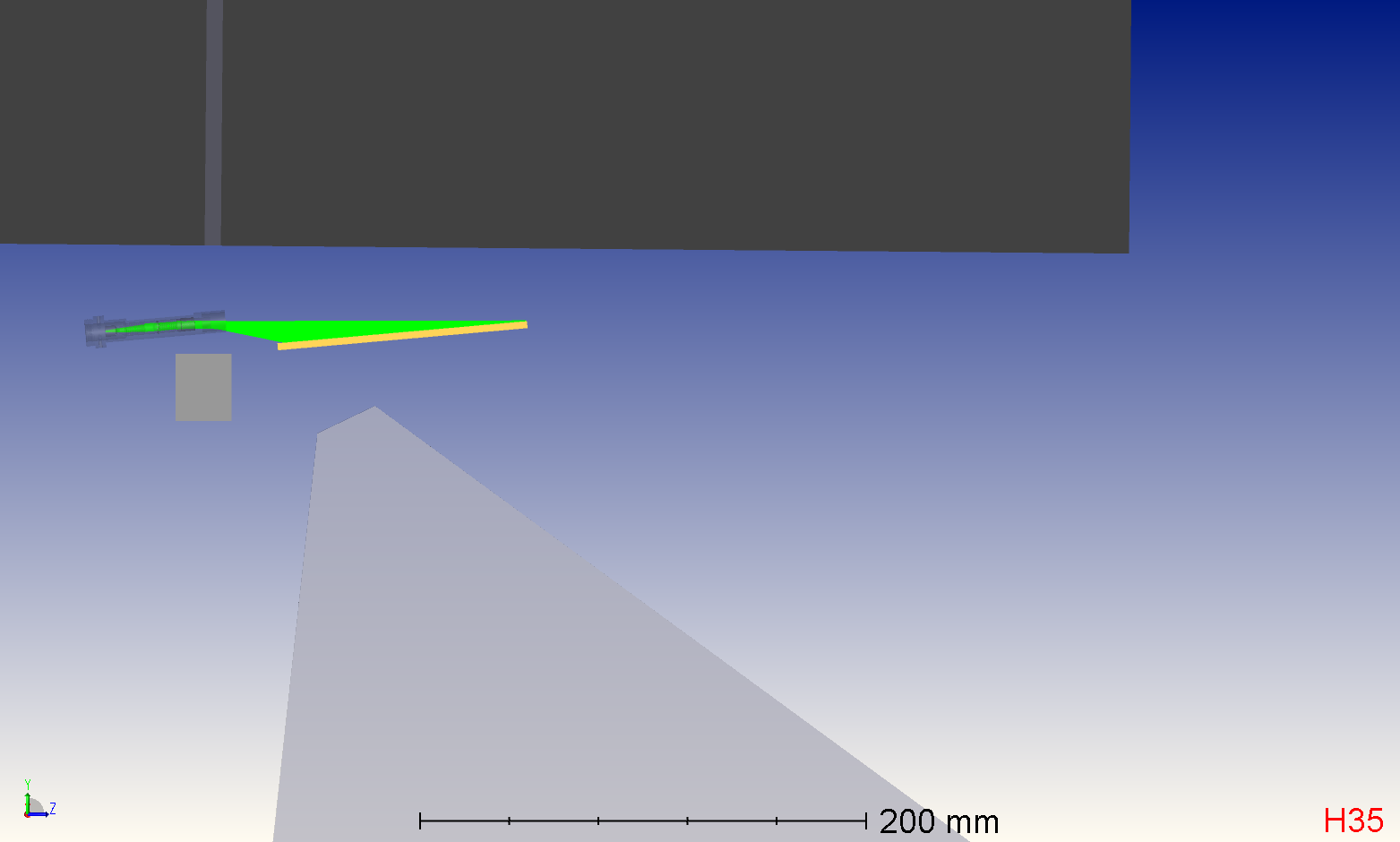}} \\
\small (b) The input optics ghost beam blocked by the baffle.
  \end{tabular}
  \caption{\textit{Zemax} rendition showing the baffle (orange) put up between the prism and the echelle grating.}   \label{fig:miniwall}
\end{figure*}

We used \textit{Zemax} Non-Sequential mode to do a ray trace using filter-strings to trace ghost beams from each element of the optical setup. Our instrument fibers have a nominal output of f/3.65, which we define such that 96$\%$ of the energy is enclosed within a cone corresponding to f/3.65. Therefore, there is a small amount of light at higher output angles. This fact can be further exacerbated due to FRD\cite{ramsey_focal_1988}, as well as misalignment of the ball lens double scrambler during the integration step\cite{halverson_efficient_2015}. This is shown in \autoref{fig:inputopticszemax}b, where the beam at high output angles is shown to be reflecting off of the outer surfaces of the lenses. In principle, this ghost reflection should be about 100 ppm after 2 bounces off of 1$\%$ AR coating (0.01$^2$ = 10$^{-4}$ = 100 ppm), however since it bypasses the echelle grating it does not get dispersed and its intensity is enhanced. To mitigate this streaking we added a vertical baffle (like a mini wall) between the prism and the grating to block the ghost beam (\autoref{fig:miniwall}). This baffle was coated with a layer of \textit{NanoLab} Singularity Black carbon nanotubes to minimize reflection from the ghost beam at glancing incidence. This baffle was sized to block the ghost from all five fibers without occulting any of the primary beam, and was designed considering the tight space constraints between the prism and the grating mounts.


\subsection{Camera ghost}\label{sec:cameraghost}

\begin{figure}[!t]
\center
\includegraphics[width=0.85\columnwidth]
{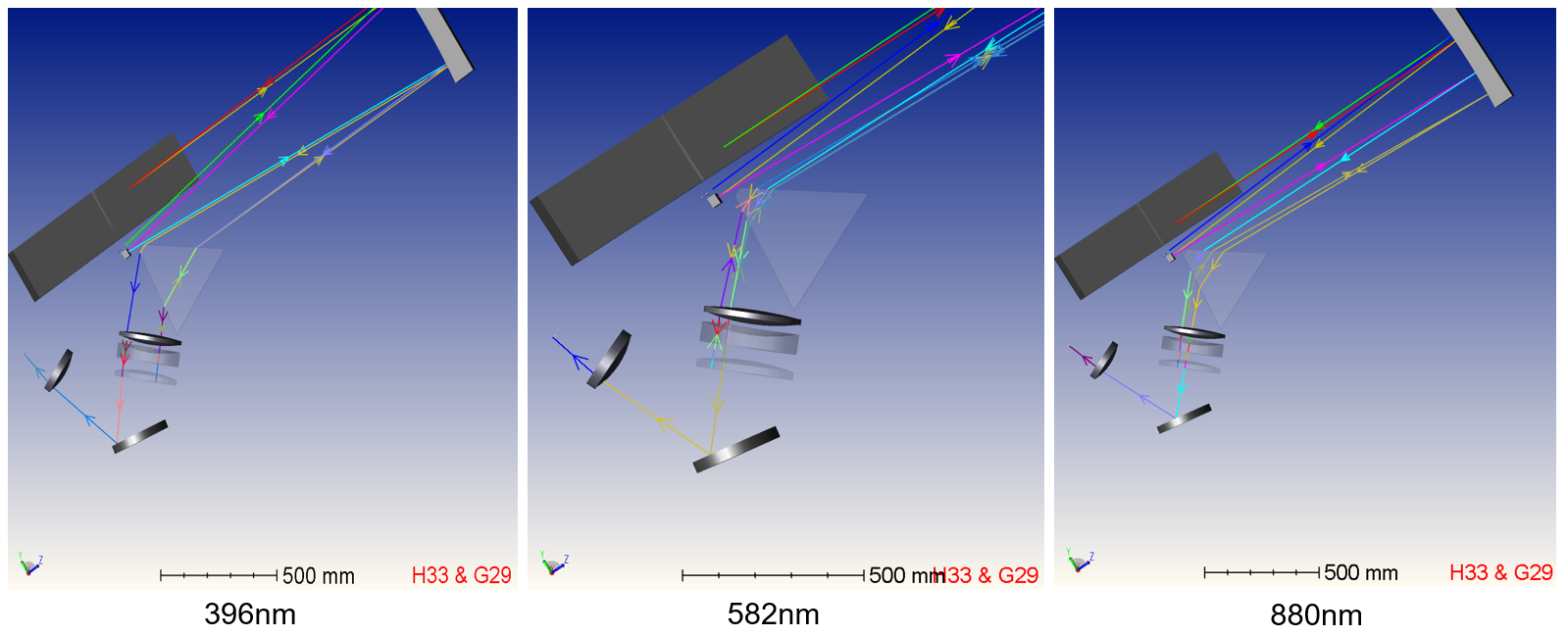}
\caption{Showing the back reflection from L3.2 for three different echelle orders, and how the diffuse ghost is formed at the same location on the detector.} \label{fig:Zemax_L3.2}
\end{figure}

The camera inside NEID consists of four refractive elements, along with a fold mirror for compactness. It has a 250 mm entrance aperture, and a $\sim 600$ mm focal length to image the entire spectral format on the detector (90 x 90 mm) \cite{schwab_design_2016}. Despite being coated with BBAR to minimize back reflections, the lens surfaces still reflect at the $\sim 1\%$ level; in particular, the back surface surface of the 3rd lens (hereafter referred to as L3.2). The L3.2 reflects back 1$\%$ of the light back to the parabola, which them forms a diffused ghost on the upper part of the focal plane (towards the red wavelengths in the cross dispersion direction) \autoref{fig:ThreeGhosts}.
This ghost from the camera lens cannot be easily mitigated without major modifications to the optical design and baffling. Since the ghost is diffused, and its contamination is mostly limited to the extreme red part of the echellogram, we decided to not make any changes to mitigate it.

\subsection{Prism Triple Bounce Ghost}\label{sec:prismtriple}
\begin{figure*}[!b]
  \centering
  \begin{tabular}[]{@{}p{0.45\textwidth}@{}}
  {
\includegraphics[width=0.42\textwidth]
{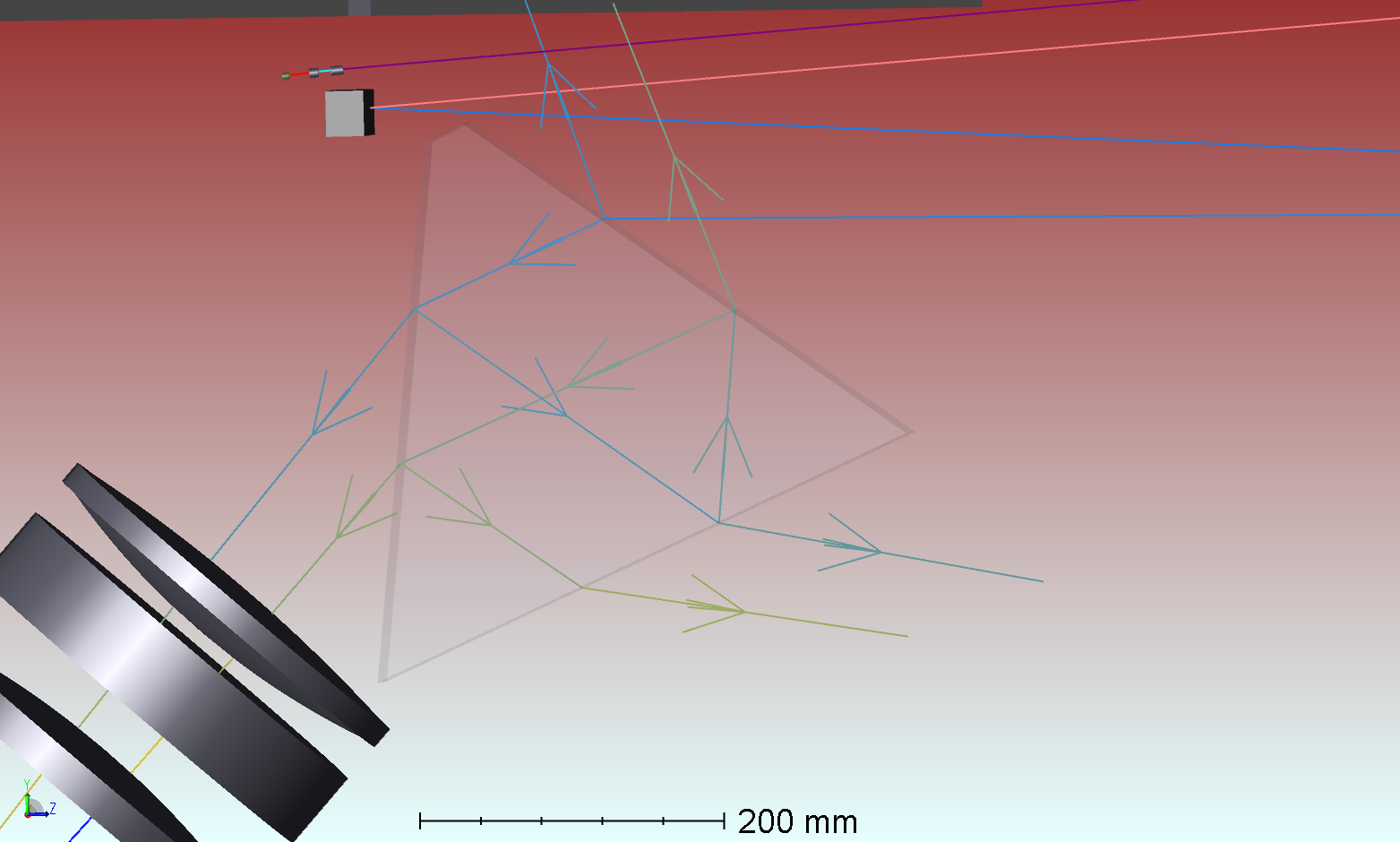}} \\\small (a) \textit{Zemax} simulation showing the prism triple bounce. 
  \end{tabular}%
  \quad
  \begin{tabular}[]{@{}p{0.43\textwidth}@{}}
 {
 \includegraphics[width=0.38\textwidth]
{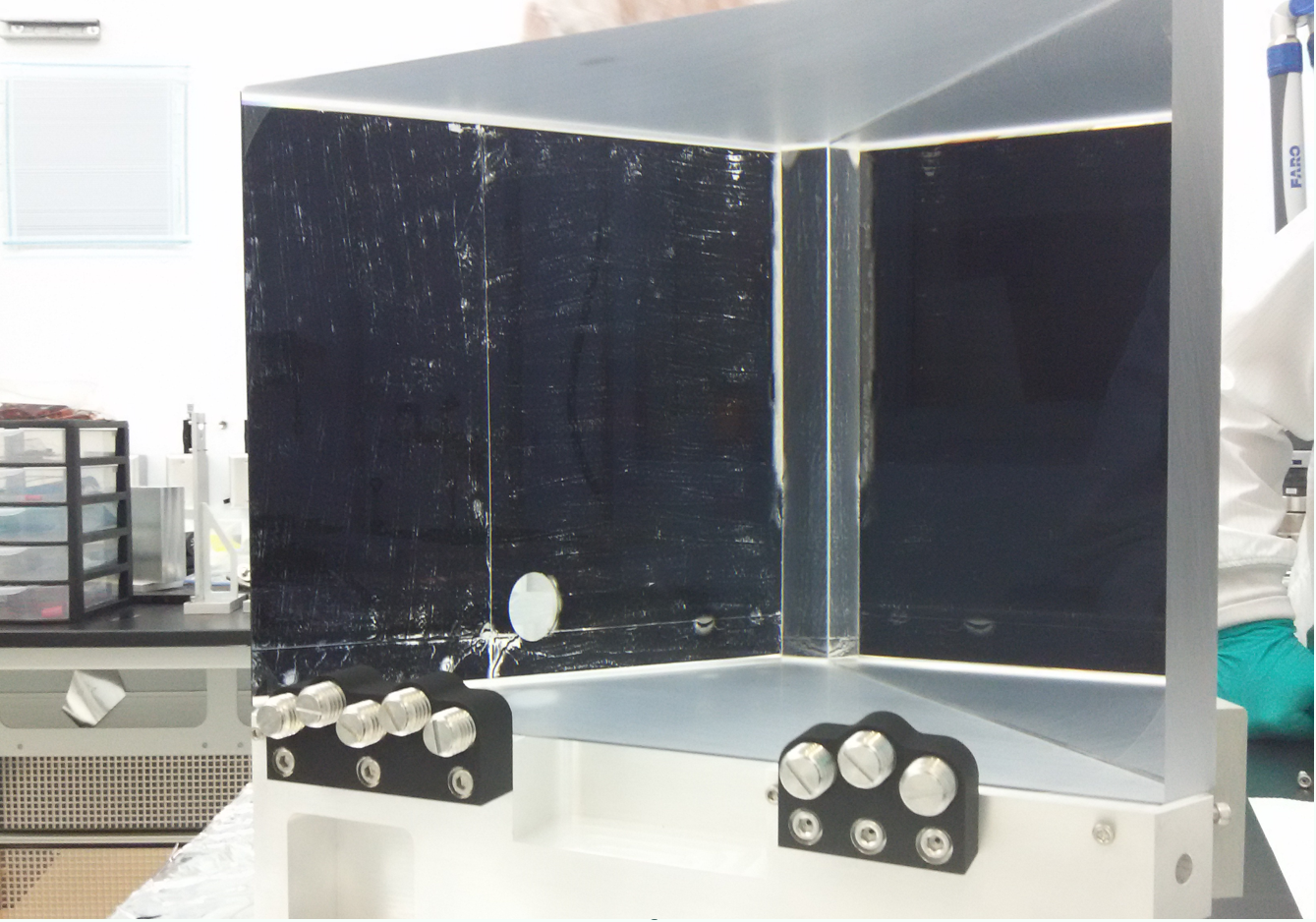}} \\
\small (b) Black Kapton applied to the base of the prism to minimize ghost reflections.
  \end{tabular}
  \caption{Prism triple bounce ghost. \textbf{a)} In this Zemax simulation the blue rays show the science beam, whereas the green ones are the ghost beam. \textbf{b)} The black kapton layer stuck to the prism minimizes Fresnel reflections and reduces the ghost by $\sim 15-20$x. }   \label{fig:prismghost}
\end{figure*}

We also notice a sharp horizontal streak (\autoref{fig:ThreeGhosts}) on the detector which is formed in the same cross dispersion position irrespective of the wavelength of the illumination. This streak is formed due to reflection from the prism surfaces despite the AR coating, particularly the base of the prism (\autoref{fig:prismghost}a)\footnote{The prism surface that is adjacent to the two optical surfaces is referred to here as the base.}. The NEID cross disperser is an \textit{Ohara} PBM2Y glass equilateral prism, with three faces that are precision MRF polished, while the two active optical surfaces are coated with an AR coating.  The ghost through the prism is not cross dispersed and hence the ghost beam from different orders (120 orders) overlap on the detector; this is because  the triple bounce is equivalent to a pair of antiparallel prisms which cancel out the cross dispersion. Furthermore, the chromaticity of the camera causes the ghost to be defocussed. For the nominal science beam, this chromaticity gets corrected by a tilt in the detector in the cross dispersion direction, however this does apply to the ghost since it is not cross dispersed. Since the base of the prism is polished, but not AR coated it experiences Fresnel reflection at about 5$\%$ amplitude. The total amplitude of this ghost should nominally be about 5 ppm (0.01$^2$ * 0.05) $\times$ 120 orders $\sim 600~ppm$ which is then defocussed to 5-10 ppm. This ghost intensity when combined from the science and calibration fibers would contaminate the science and calibration traces that overlap with this beam.

\begin{figure*}[!t]
  \centering
  \begin{tabular}[]{@{}p{0.45\textwidth}@{}}
  {
\includegraphics[width=0.42\textwidth]
{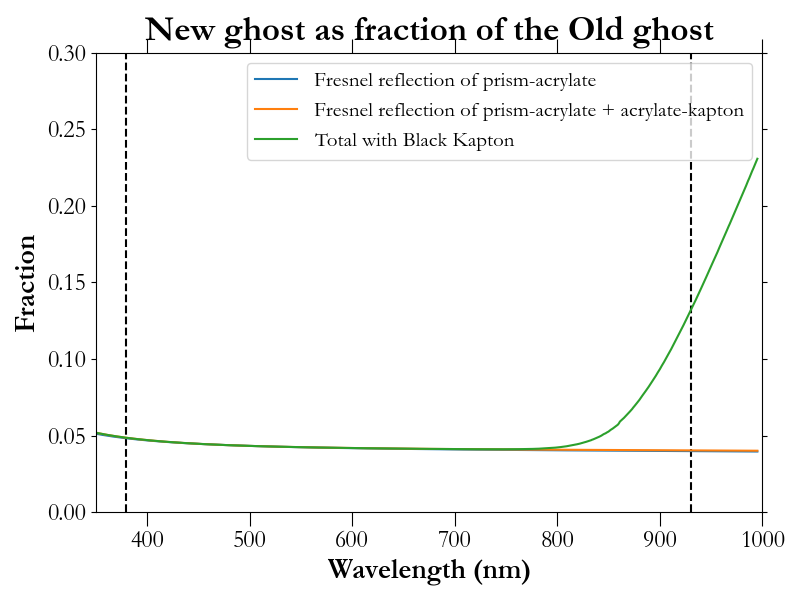}} \\\small (a) Ghost fraction
  \end{tabular}%
  \quad
  \begin{tabular}[]{@{}p{0.45\textwidth}@{}}
 {
 \includegraphics[width=0.42\textwidth]
{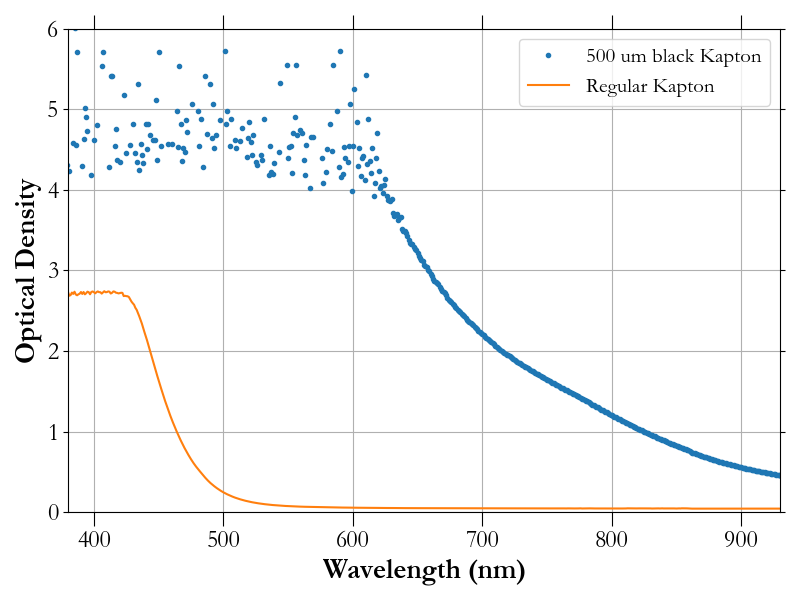}} \\
\small (b) The Optical Density (OD) for Kapton, where the transmission (T) = 10$^{-{\rm{OD}}}$
  \end{tabular}
  \caption{\textbf{a)} Theoretical predictions for the prism triple bounce ghost after applying the Kapton, as compared to the original ghost. The majority of the new ghost beam is from the Fresnel reflection between the prism-acrylate interface due to the index mismatch (1.6 vs 1.48) and is shown in blue, while the Fresnel reflection from the acrylate-Kapton layer is negligible. Therefore the blue and orange curves pretty much lie on top of each other. The total contribution (green) includes the double pass through the Kapton layer. The two vertical dashed lines mark the NEID bandpass. \textbf{b)} The absorbance of the black Kapton film showing how it absorbs most of the blue light, but is more transparent in the red. The scatter in the blue is due to measurement uncertainty at higher ODs. For comparison we also show the absorbance of regular Kapton.}   \label{fig:prismghostfraction}
\end{figure*}

We reduced the magnitude of this contamination by mitigating the Fresnel reflection from the prism base (n $\sim 1.6$). We use an acrylate\footnote{\textit{Thorlabs} OCA8146-3} double sided adhesive of refractive index (n $\sim 1.47$) to stick a 500 $\mu$m black Kapton\footnote{DuPont$\texttrademark$ Kapton$^{\textcopyright}$} film\footnote{\textit{McMaster} 7161T21} (n $\sim 1.48$) on the prism face. The Kapton+acrylate layer reduces the Fresnel reflection from the prism-air interface ($\sim 5\%$) by about 20x (\autoref{fig:prismghost}b). The new ghost reflection from the prism base primarily consists of the  Fresnel reflection from the prism-acrylate interface  and that from the acrylate+Kapton interface (\autoref{fig:prismghostfraction}). In addition, a small fraction transmits through the Kapton and partially reflects back due to Fresnel reflection from the Kapton - air interface. This is primarily the case for the redder wavelengths, since the Kapton absorbs most of the bluer wavelengths. Overall this represents about a 20x improvement in the ghost across most of the NEID bandpass, rendering the ghost largely undetectable during regular science operations.

\begin{figure}[!b]
\center
\includegraphics[width=0.4\columnwidth]
{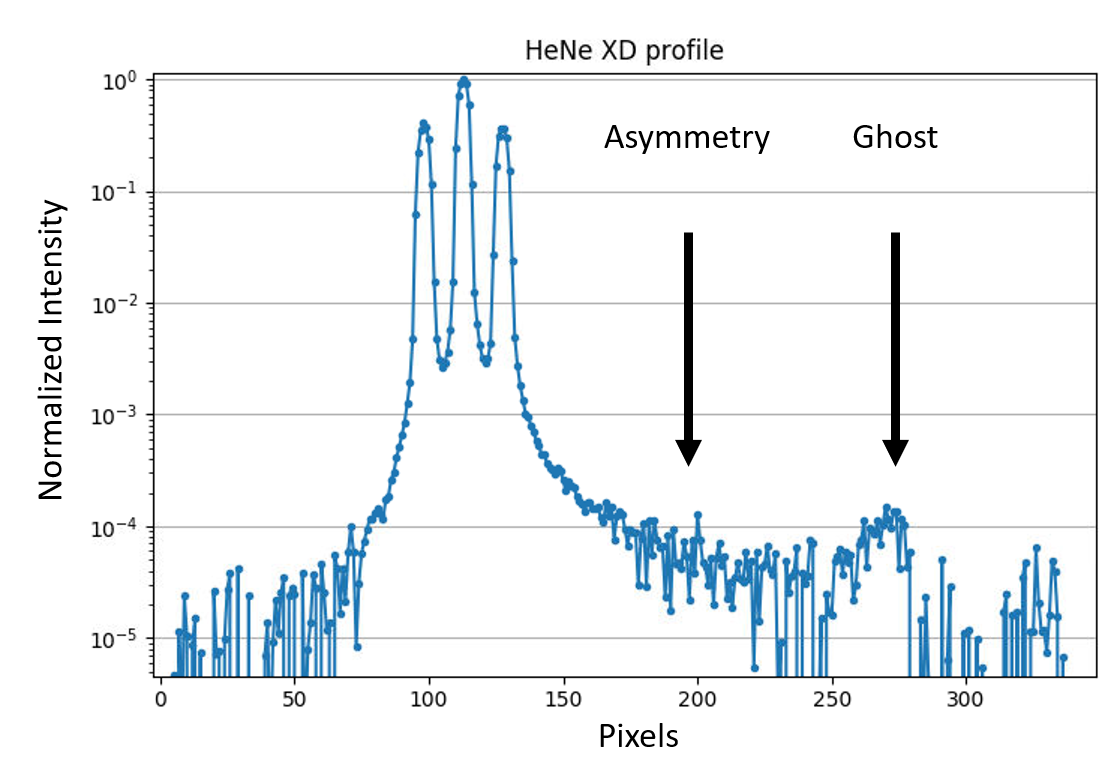}
\caption{A cross dispersion profile of the 3 NEID fibers when illuminated by a Helium - Neon laser at 632.8 nm. The wavelength increases in the same direction as pixels; the wing asymmetry and ghost are marked with black arrows.} \label{fig:redwingghost}
\end{figure}

\subsection{Red Wing Ghost}
We noticed an asymmetry in the PSF with an enhanced red wing and a ghost at the end (Figure \ref{fig:redwingghost}). This ghost was seen to be chromatically dispersed on the detector, and offset by about 120 pixels\footnote{1 NEID pixel = 10 $\mu$m} in the cross dispersion direction direction from the primary fiber trace. Since the dispersion of the ghost was seen to be similar to the primary trace, we concluded that the ghost was formed on or after the grating. This ghost if uncorrected would significantly increase the inter-order contamination. The calibration spectra could contaminate the PSF centroids for the stellar spectrum, while the stellar spectra would contaminate the calibration spectra, reducing the accuracy of instrumental drift measurements.

\begin{figure}[!t]
\center
\includegraphics[width=0.85\columnwidth]
{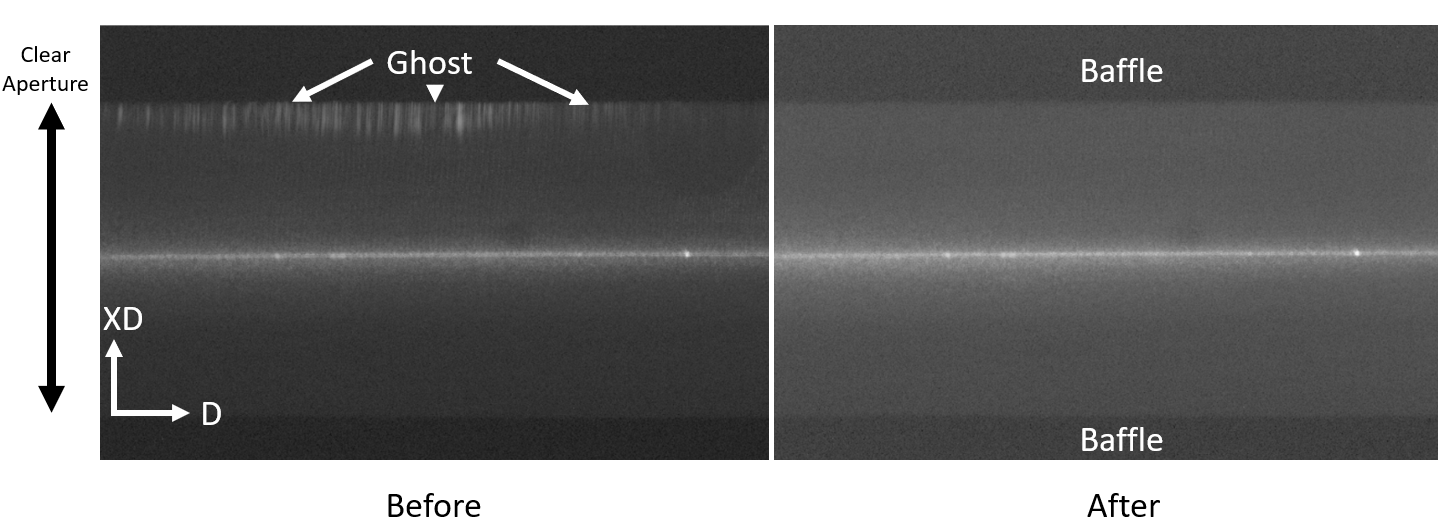}
\caption{An image of the trace formed on the fold mirror when illuminated by an off-the-shelf diode laser. The dispersion (D) and cross dispersion (XD) directions are marked with the white arrows. The extent of the clear aperture on the fold mirror is seen, along with the extent of the baffling. In the left image (before), the ghost can be seen on one side of the PSF, which matches the profile seen on the FPA. The right image (after) shows how the mitigation strategy gets rid of the ghost.} \label{fig:M2Redwing}
\end{figure}

To diagnose and mitigate the ghost, we imaged the intermediate focus formed at the fold mirror (M2) using a \textit{Lodestar} detector, as a proxy for the image formed on the FPA (\autoref{fig:M2Redwing}). Using this as a diagnostic, we tested various hypotheses for the possible progenitor of this red wing ghost.
We considered the possibility that the ghost could be formed from a back reflection off of the fold mirror substrate, since the mirror has a reflectivity of $\sim 98\%$. \textit{Zemax} simulations reveal that the second bounce from the (second surface) fold mirror substrate would form an image with a similar offset on the detector. However a double pass through a Zerodur substrate of thickness 25 mm attenuates the blue wavelengths substantially more than the red; which does not match the chromatic profile seen for the ghost. To mitigate a possible reflection from the inside of the aluminium mount, we applied a layer of Kapton film coated with carbon nanotubes to this surface.

\begin{figure*}[!t]
  \centering
  \begin{tabular}[]{@{}p{0.4\textwidth}@{}}
  {
\includegraphics[height=0.42\textwidth]
{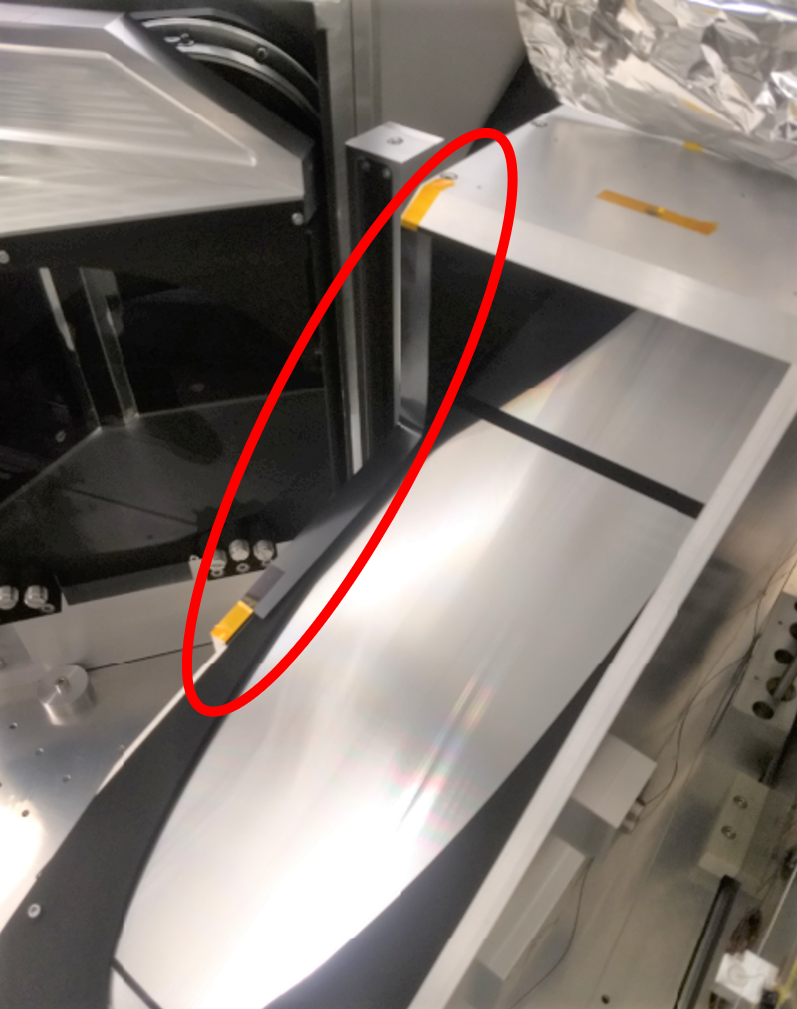}} \\\small 
  \end{tabular}%
  \quad
  \begin{tabular}[]{@{}p{0.45\textwidth}@{}}
 {
 \includegraphics[height=0.42\textwidth]
{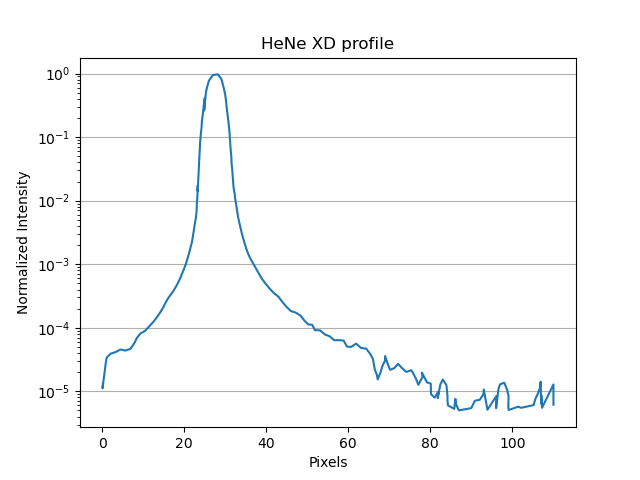}} \\
\small 
  \end{tabular}
  \caption{\textbf{a)} The temporary Kapton baffle used for testing (circled in red over here), installed over the grating baffle, blocks a tiny sliver of the edge of the grating baffle that mitigates the red wing ghost. \textbf{b)} Cross dispersion profile of the science fiber illuminated by a HeNe laser after mitigation of the red wing ghost by blocking reflection off the edge of the grating baffle.}   \label{fig:RedWingBaffleKapton}
\end{figure*}

During the NEID optical design and simulation step we noticed that the echelle grating mount interfered with the diffracted return beam from the collimating M1 mirror to the fold mirror. This obstruction preferentially blocked $\sim$ 5$\%$ light for the redder wavelengths of each order ($\lambda >$  800 nm). We fixed this by modifying the design of the mount and shifting the grating away from the collimating mirror (since it is illuminated in collimated space), and away from the fold mirror. This movement was within the clear aperture of the grating, and hence did not affect the optical performance of the instrument. We considered if a grazing reflection of the pupil off the baffle edge could be the progenitor of the red wing ghost, and tested this by adding a temporary Kapton tape obscuration in front of the baffle edge (\autoref{fig:RedWingBaffleKapton}a) while imaging the ghost on the fold mirror. This Kapton strip was coated with carbon nanotubes to prevent scattered light from the highly oblique incident ($i \sim 76^{\circ}$) light. As can be seen in the `Before` and `After` images in \autoref{fig:M2Redwing}, this Kapton obscuration effectively blocks the ghost. When we took a similar HeNe laser exposure on the NEID detector, we see the same mitigation of the ghost in the cross dispersion profile of the science fiber (\autoref{fig:RedWingBaffleKapton}b). Therefore this arrangement with the baffle was finalized and added to the instrument setup.

\section{Conclusion}
In this manuscript we discuss the sources of stray light contamination in the NEID spectrograph, and the steps taken to ameliorate their spectral impact. We share the lessons learnt and the tools we used during this process, to help the design and integration of similar instruments. In particular, optical design and simulation tools such as \textit{Zemax} Non-Sequential, \textit{LensMechanix} and \textit{SolidWorks} are extremely useful to explore the interplay between the optical components and their mounts. 

The ultimate goal for the RV method of planet detection and mass measurement is to have instruments that are precise enough to measure the Doppler signal from an Earth analogue. This requires managing each of the contributing terms in the error budget\cite{halverson_comprehensive_2016}, and their impact on the RV measurement precision of the instrument. We thereby describe the process followed to reduce the impact of stray light contamination for NEID, and hope that these investigations will be useful for other instrument teams.


\acknowledgments 
 
NEID is funded by JPL under contract 1547612.  This work was partially supported by funding from  the  Center  for  Exoplanets  and  Habitable  Worlds.   The  Center  for  Exoplanets  and  Habit-able  Worlds  is  supported  by  the  Pennsylvania  State  University,  the  Eberly  College  of  Science,and the Pennsylvania Space Grant Consortium.  We  acknowledge  support from NSF grants AST-1006676, AST-1126413, AST-1310885, the NASA Astrobiology Institute(NAI; NNA09DA76A), and the Penn State Astrobiology Research Center.
\bibliography{MyLibrary} 
\bibliographystyle{spiebib} 

\end{document}